%% file: gopalswamy.tex
\documentclass{svmult}       
\usepackage{epsfig,graphicx}   
\usepackage[bottom]{footmisc}  
\usepackage{natbib,url}      
\usepackage{bm}              
\usepackage{amssymb}         
\bibpunct{(}{)}{;}{a}{}{,}   
\def\cite#1{\citealp{#1}}    
\def\authorindex#1{}         

\begin{document}\newcount\preprintheader\preprintheader=1 

\input{rr-assp-defs}


\title*{Coronal Mass Ejections from Sunspot and non-Sunspot Regions}


\author{N. Gopalswamy\inst{1}
        \and
        S. Akiyama\inst{1,2}
        \and
        S. Yashiro\inst{1,2,3}
        \and
        P. M\"akel\"a\inst{1,2}}

\authorindex{Gopalswamy, N.}
\authorindex{Akiyama, S.}
\authorindex{Yashiro, S.}
\authorindex{M\"akel\"a, P.}

\institute{NASA Goddard Space Flight Center, Greenbelt, USA
           \and 
           The Catholic University of America, Washington, USA
           \and
           Interferometrics, Herndon, USA}


\maketitle

\setcounter{footnote}{0}  

\begin{abstract} 
  Coronal mass ejections (CMEs) originate from closed magnetic field
  regions on the Sun, which are active regions and quiescent filament
  regions.  The energetic populations such as halo CMEs, CMEs
  associated with magnetic clouds, geoeffective CMEs, CMEs associated
  with solar energetic particles and interplanetary type II radio
  bursts, and shock-driving CMEs have been found to originate from 
  sunspot regions. The CME and flare occurrence rates are found to be
  correlated with the sunspot number, but the correlations are
  significantly weaker during the maximum phase compared to the rise
  and declining phases.  We suggest that the weaker correlation results
  from high-latitude CMEs from the polar crown filament regions
  that are not related to sunspots.
\end{abstract}

\section{Introduction}      \label{gopalswamy-sec:introduction}

Coronal mass ejections (CMEs) are the most energetic phenomena in the
solar atmosphere and represent the conversion of stored magnetic
energy into plasma kinetic energy and flare thermal energy.  The
transient nature of CMEs contrasts them from the solar wind, which is
a quasi steady plasma flow.  Once ejected, CMEs travel through the
solar wind and interact with it, often setting up fast-mode MHD
shocks, which in turn accelerate charged particles to very high
energies. CMEs often propagate far into the interplanetary (IP) medium
impacting planetary atmospheres and even the termination shock of the
heliosphere.  The magnetic fields embedded in CMEs can merge with
Earth's magnetic field resulting in intense geomagnetic storms, which
have serious consequences throughout the geospace and even for life on
Earth.  Thus, CMEs represent magnetic coupling at various locations in
the heliosphere.  Active regions on the Sun, containing sunspots and
plages, are the primary sources of CMEs.  Closed magnetic field
regions such as quiescent filament regions also cause CMEs.  These
secondary source regions can occur at all latitudes, but during the
solar maximum, they occur prominently at high latitudes, where
sunspots are not found. This paper summarizes the properties of CMEs
as an indicator of solar activity in comparison with the sunspot
number.

\begin{figure}  [b]
  \centering
  \includegraphics[width=\textwidth]{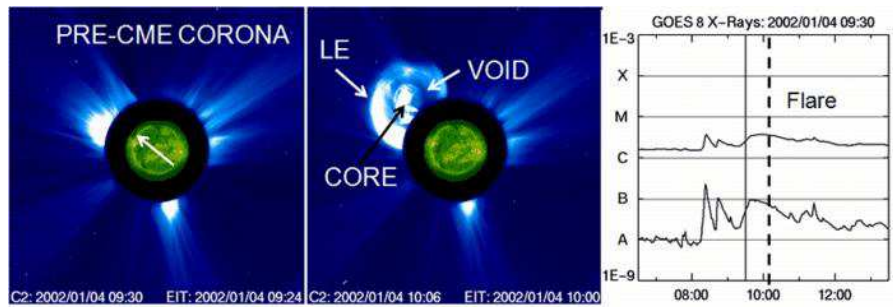}
  \caption[]{\label{gopalswamy-fig:example}  
%
  Example of a CME originating from near the northeast limb of the Sun
  (pointed by arrow) as a distinct structure into the pre-CME
  corona. The CME roughly fills the northeast quadrant of the Sun. The
  three primary structures of the CME, viz., the leading edge (LE),
  which is curved like a loop in 2D projection, the dark void, and the
  structured prominence core are indicated by arrows. The plot to the
  right shows the GOES soft X-ray flare associated with the CME.  The
  vertical solid line marks the LASCO frame at 09:30 UT (pre-CME
  corona) and the dashed line marks the frame with the CME at 10:06
  UT.
}\end{figure}

\begin{figure}  
  \centering
  \includegraphics[width=0.8\textwidth]{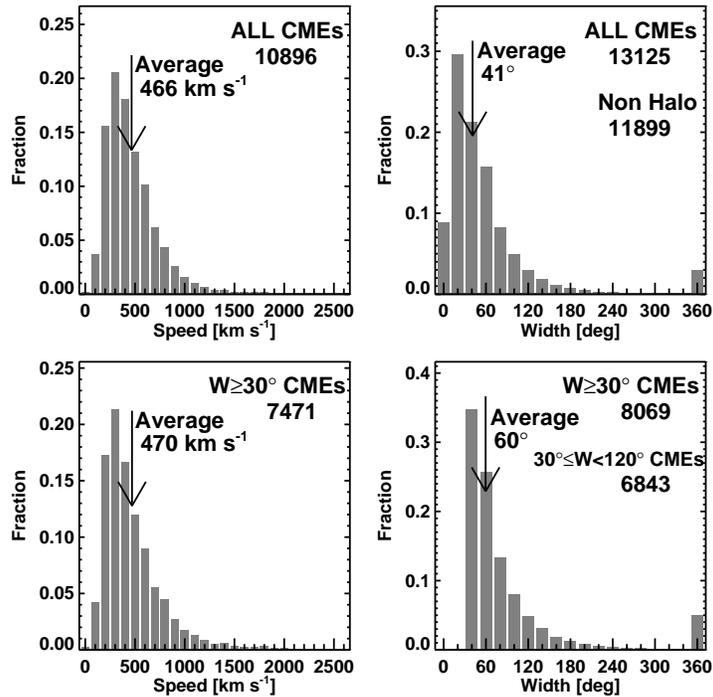}
  \caption[]{\label{gopalswamy-fig:spdwd}  
%
Speed and width distributions of all CMEs (top) and non-narrow CMEs
($W\ge30^\circ$; bottom). The average width of non-narrow CMEs is
calculated using only those CMEs with $W\ge30^\circ$.
}\end{figure}

\section{Summary of CME properties} \label{gopalswamy-sec:cme}

\begin{figure}  
  \centering
  \includegraphics[width=\textwidth]{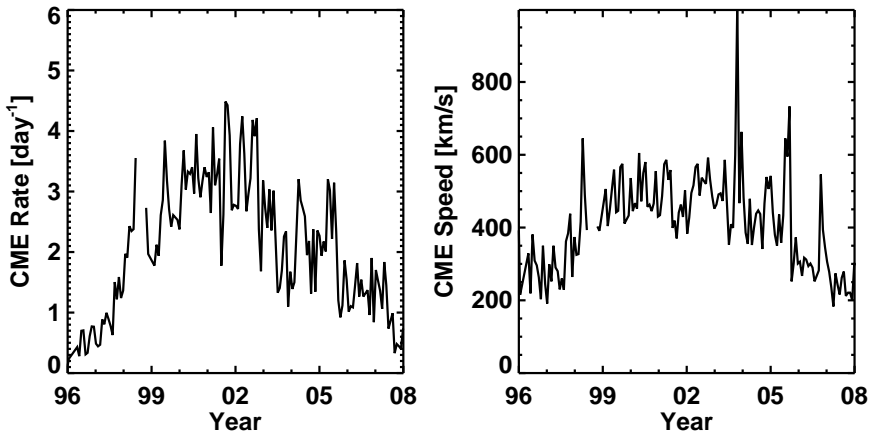}
  \caption[]{\label{gopalswamy-fig:cmerate}  
%
  The daily CME rate (for CMEs with $W\ge30^\circ$) and the mean CME
  speed plotted as a function of time showing the solar cycle
  variation. The occasional spikes are due to super-active regions.
}\end{figure}

Figure~\ref{gopalswamy-fig:example} illustrates a CME as a large-scale
structure moving in the corona and the associated soft X-ray
flare. The CME observations were made by the Solar and Heliospheric
Observatory (SOHO) Mission's Large Angle and Spectrometric Coronagraph
(LASCO). The CME is clearly an inhomogeneous structure with a well
defined leading edge (LE) followed by a dark void and finally an
irregular bright core.  The core is nothing but an eruptive prominence
normally observed in H$\alpha$ or microwaves, but here observed in the
photospheric light Thomson-scattered by the prominence. Prominence
eruptions and flares have been known for a long time before the
discovery of CMEs in the early 1970s. Several coronagraphs have
operated since then and have accumulated a wealth of information on
the properties of CMEs (see, e.g., Hundhausen, 1993; Gopalswamy 2004;
Kahler, 2006). Here we summarize the statistical properties of CMEs
detected by SOHO/LASCO and compiled in a catalog (Gopalswamy et~al.,
2009c):

\begin{itemize} \itemsep=1ex 

  \item The CME speed is obtained by tracking the leading edge until
        it reaches the edge of the LASCO field of view (FOV, extending
        to about 32~R$_\odot$). Some CMEs become faint before reaching
        the edge of the FOV and others farther. Therefore, the CME
        speed we quote here is an average value within the LASCO FOV.
        Since the height-time measurements are made in the sky plane,
        the speed is a lower limit. Figure~\ref{gopalswamy-fig:spdwd}
        shows that the speed varies over two orders of magnitude from
        20~km/s to more than 3000~km/s, with an average value of
        466~km/s.

  \item The CME angular width is measured as the position angle extent
        of the CME in the sky plane. Figure~\ref{gopalswamy-fig:spdwd}
        shows the width distribution for all CMEs and for CMEs with
        width $>$30$^\circ$. The narrow CMEs ($W<30^\circ$) were
        excluded because the manual detection of such CMEs is highly
        subjective (Yashiro et~al., 2008b). The apparent width ranges
        from $<$5$^\circ$ to 360$^\circ$ with an average value of
        41$^\circ$ (60$^\circ$ when CMEs wider than 30$^\circ$
        are considered).  There is actually a correlation between CME
        speed ($V$~km/s) and width ($W$ in degrees) indicating that
        faster CMEs are generally wider: $V = 360 + 3.64\,W$ (Gopalswamy
        et~al., 2009a).

  \item CMEs with the above-average speeds decelerate due to coronal
        drag, while those with speeds well below the average
        accelerate. CMEs with speeds close to the average speed do not
        have observable acceleration.

  \item The CME mass ranges from 10$^{12}$~g to $>$10$^{16}$~g
        with an average value of 10$^{14}$~g.  Wider CMEs
        generally have a greater mass content ($M$): $\log M = 12.6 +
        1.3\log W$ (Gopalswamy et~al. 2005).  From the observed mass
        and speed, one can see that the kinetic energy ranges from
        10$^{27}$~erg to $>$10$^{33}$~erg, with an average value
        of $5.4\times10^{29}$~erg.

  \item The daily CME rate averaged over Carrington rotation periods
        ranges from $<0.5$ (solar minimum) to $>6$ (solar
        maximum). The average speed increases from about 250~km/s
        during solar minimum to $>550$~km/s during solar maximum (see
        Fig.~\ref{gopalswamy-fig:cmerate}).

  \item CMEs moving faster than the coronal magnetosonic speed drive
        shocks, which accelerate solar energetic particles (SEPs) to
        GeV energies. The shocks also accelerate electrons, which
        produce nonthermal radio emission (type II radio bursts)
        throughout the inner heliosphere. 

  \item The CME eruption is accompanied by solar flares whose
        intensity in soft X-rays is correlated with the CME kinetic
        energy (Hundhausen, 1997; Yashiro \& Gopalswamy, 2009). 

  \item There is a close temporal and spatial connection between CMEs
        and flares: CMEs move radially away from the eruption region,
        except for small deviations that depend on the phase of the
        solar cycle (Yashiro et~al., 2008a). However, more than half of
        the flares are not associated with CMEs. 

  \item CMEs are comprised of multithermal plasmas containing coronal
        material at a temperature of a few times $10^6$~K and
        prominence material at about $8000$~K in the core. In-situ
        observations show high charge states within the CME, confirming
        the high temperature in the eruption region due to the flare
        process associated with the CME (Reinard, 2008).

  \item CMEs originate from closed field regions on the Sun, which are
        active regions, filament regions, and transequatorial
        interconnecting regions. 

  \item Some energetic CMEs move as coherent structures in the
        heliosphere all the way to the edge of the solar system. 

  \item Theory and IP observations suggest that CMEs
        contain a fundamental flux rope structure identified with the
        void structure in Fig.~\ref{gopalswamy-fig:example} (see e.g.,
        Gopalswamy et~al., 2006; Amari \& Aly, 2009). The prominence core is
        thought to be located at the bottom of this flux rope. The
        frontal structure is the material piled up at the leading edge
        of the flux rope.  

  \item CMEs are often associated with EUV waves, which may be
        fast mode shocks when the CME is fast enough (Neupert, 1989;
        Thompson et~al., 1998).

  \item Coronal dimmings are often observed as compact regions located
        on either side of the photospheric neutral line, which are
        thought to be the feet of the erupting flux rope (Webb
        et~al. 2000).

\end{itemize}

\begin{figure}  
  \centering
  \includegraphics[width=\textwidth]{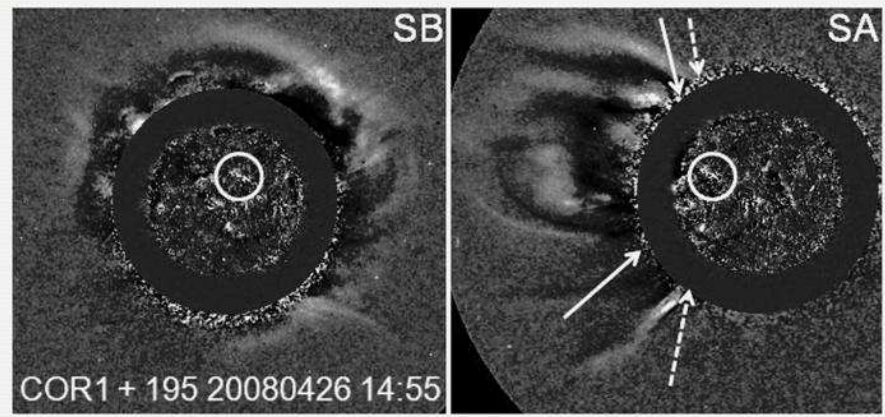}
  \caption[]{\label{gopalswamy-fig:halo}  
%
  Two images of the 2008 April 26 CME observed simultaneously by the
  two STEREO spacecraft separated by about $50^\circ$.  Clearly the CME
  appears as a halo in the view of the Behind spacecraft (SB), which
  trailed Earth by about $25^\circ$.  On the other hand, the CME appears
  as a normal CME in the view of the Ahead spacecraft (SA) which was
  ahead of Earth by about $25^\circ$.  In the SA view, the CME is
  confined to the angular extent bounded by the two solid arrows. The
  dashed arrows indicate the disturbances surrounding the CME, which
  are likely to be shocks for fast CMEs. In the SB view, it is
  difficult to see the inner and outer structures, so what is seen is
  likely to be the disturbances surrounding the CME. The solar source
  of the CME is shown encircled in the difference image taken by
  STEREO's Extreme Ultraviolet Imager (EUVI). The elongated feature is
  the post eruption arcade, which is normally taken as the solar
  source of the CME. The halo CME is roughly symmetric around the
  solar source in SB view, while the CME is mostly to the east in the
  SA view.  Thus a top view produces a halo CME, while a broadside
  view produces a normal CME. Occasionally, the CME can become a halo
  even in the broadside view; such CMEs are known as ``limb halos''
  because the disturbances can be seen on the limb opposite to the
  solar source. Such limb halos are generally the fastest events.
}\end{figure}

\begin{figure}  
  \centering
  \includegraphics[width=\textwidth]{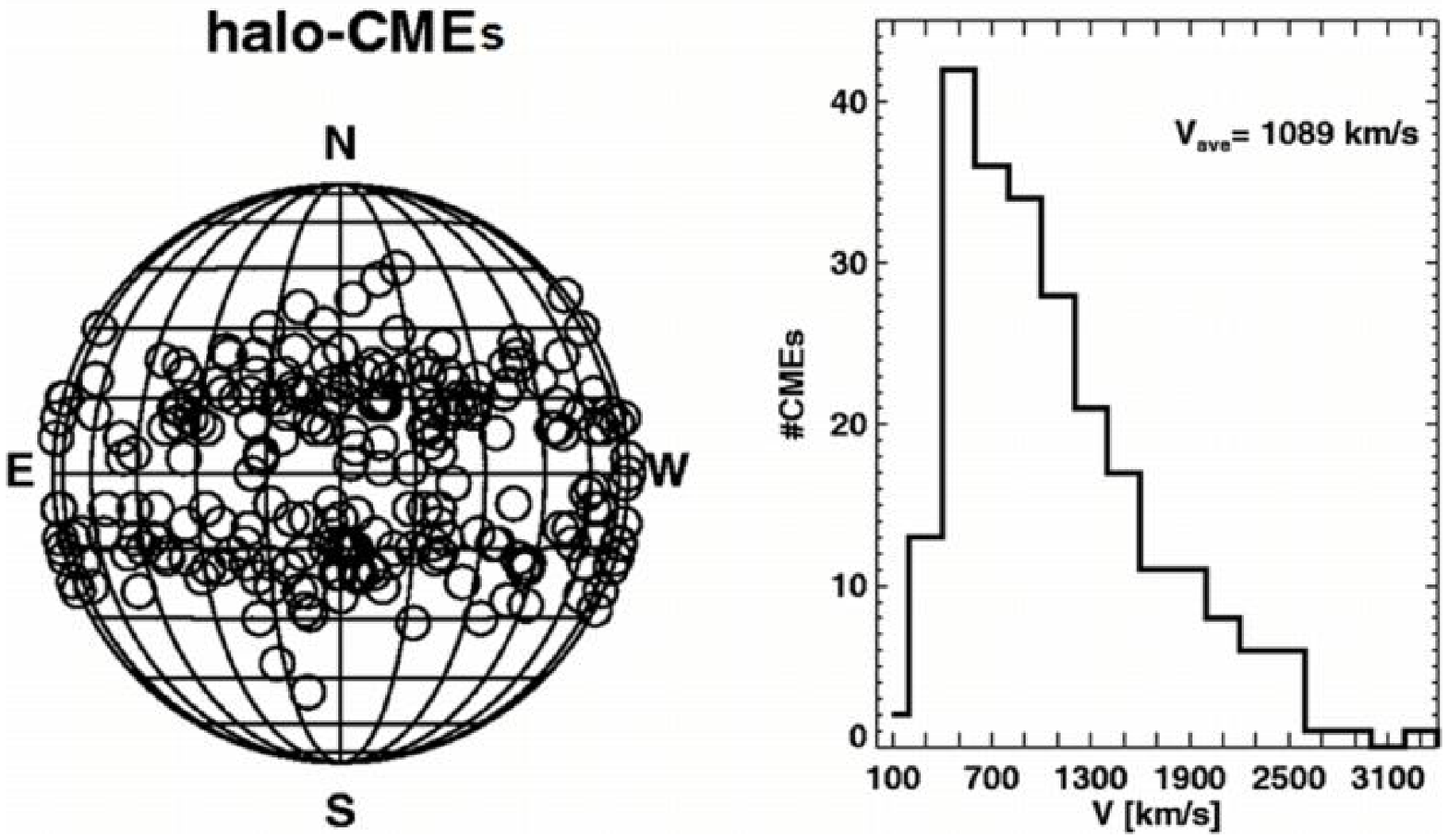}
  \caption[]{\label{gopalswamy-fig:halo-source}  
%
  Source locations ({\em left\/}) and the speed distribution ({\em right\/}) of halo
  CMEs recorded by SOHO during 1996-2007. Note the heavy concentration
  of the sources close to the central meridian. The average speed of
  halos ($V_{ave}$) is more than two times the average speed of the
  general population.
}\end{figure}

\section{Special population of CMEs} \label{gopalswamy-sec:special}

In this section, we consider several subsets of CMEs that have
significant consequences in the heliosphere: halo CMEs, SEP-producing
CMEs, CMEs associated with IP type II radio bursts, CMEs associated
with shocks detected in situ, CMEs detected at 1 AU as magnetic clouds
and non-cloud ICMEs.

\subsection{Halo CMEs}

CMEs appearing to surround the occulting disk in coronagraphic images
are known as halo CMEs (Howard et~al., 1982). Halo CMEs are like any
other CMEs except that they move predominantly toward or away from the
observer.  Figure~\ref{gopalswamy-fig:halo} illustrates this
geometrical effect: the same CME is observed from two viewpoints by
the Solar Terrestrial Relations Observatory (STEREO) coronagraphs
separated by an angle of $50^\circ$.  In the SOHO data $3.6$\% of all
CMEs were found to be full halos, while CMEs with $W\ge120^\circ$
account for $11$\% (Gopalswamy, 2004). Coronagraph occulting disks
block the solar disk and the inner corona, so observations in other
wavelengths such as H$\alpha$, microwave, X-ray or EUV are needed to
determine whether a CME is frontsided or not.
Fig.~\ref{gopalswamy-fig:halo-source} shows the source locations of
halo CMEs defined as the heliographic coordinates of the associated
flares. Clearly, most of the sources are concentrated near the central
meridian. Halos within a central meridian distance (CMD) of $45^\circ$
are known as disk halos, while those with
$45^\circ<\rm{CMD}\le90^\circ$ are known as limb halos.  Halos
occurring close to the central meridian appear as symmetric halos in
the coronagraph images. Halos originating from close to the limb
appear asymmetric (see Gopalswamy et~al., 2003a). The outermost
structures seen in most of the halos are likely to be the disturbances
(shocks) surrounding the main body of the CME (Sheeley et~al., 2000).
When the halo CMEs erupt on the visible face of the Sun, they are
highly likely to impact Earth and cause intense geomagnetic
storms. About 70\% of the halos have been found to be geoeffective
($Dst\le-50$~nT).  Disk halos are more geoeffective by their direct
impact on Earth, while the limb halos are less geoeffective because of
the glancing impact they deliver. More details on halo CMEs can be
found in Gopalswamy et~al. (2007).

The speed distribution in Fig.~\ref{gopalswamy-fig:halo-source} shows
that the halos are high-speed CMEs (the average speed is
about 1090~km/s, more than two times the average speed of all
CMEs). The halo CMEs are also associated with flares of greater X-ray
importance (Gopalswamy et~al., 2007).  Halo CMEs observed from a
single spacecraft have no width information. However, higher-speed
CMEs have been found to be wider from a set of CMEs erupting close to
the limb (Gopalswamy et~al., 2009a). Furthermore, wider CMEs are more
massive (Gopalswamy et~al., 2005). Therefore, one can conclude that
halos are more energetic on the average.  This is the reason that a
large fraction of halos are important for space weather as they drive
shocks that produce SEPs and produce geomagnetic storms.

\subsection{CMEs resulting in ICMEs}

CMEs in the IP medium are known as IP CMEs or ICMEs for short. ICMEs
are generally inferred from single point observations from the solar
wind plasma, magnetic field and composition signatures as the ICME
blows past the observing spacecraft (see Gosling, 1996 for a
discussion on the identification of ICMEs).  ICMEs can be traced back
to the frontside of the Sun by looking at 5 days of solar observations
preceding the ICME arrival at Earth. When ICMEs have a flux rope
structure with depressed proton temperature compared to the pre-ICME
solar wind, they are known as magnetic clouds (MCs) or IP flux ropes
(Burlaga et~al., 1981). CMEs resulting in MCs generally originate
close to the disk center. CMEs resulting in non-cloud ICMEs generally
originate at CMD $>30^\circ$, but there are many exceptions.
Shock-driving ICMEs are easier to identify because shocks are
well-defined discontinuities observed in the solar wind. Sometimes, IP
shocks are observed without a discernible ICME.  In these cases the
CMEs are ejected at large angles to the Sun-Earth line, so the ICME
part misses Earth. Thus as one goes from the disk center to the limb,
one encounters solar sources of MCs close to the disk center, those of
non-clouds ICMEs, and finally shocks without drivers (Gopalswamy,
2006). This suggests the IP manifestation of CMEs depends on the
observer--Sun--CME angle.  Assuming that CMEs reaching 1~AU have an
average width of $60^\circ$, one can see that about one third of
such CMEs should be MCs and the remaining two-thirds should be
non-cloud ICMEs (excluding the pure shock cases). This is generally
the case on the average (e.g., Burlaga, 1995), but there are notable
exceptions: (i) the fraction of MCs is very high during the rise phase
of the solar cycle compared to the maximum phase (Riley et~al., 2006),
(ii) there are many non-cloud ICME sources close to the disk center,
(iii) some pure-shock sources are close to the disk center during the
declining phase. These exceptions can be explained as the effect of
external influences (Gopalswamy et~al., 2009b).

ICMEs reaching Earth are highly likely to cause magnetic storms,
provided they contain south-pointing magnetic field either in the ICME
portion or in the sheath portion or both. In fact about 90\% of the
large geomagnetic storms are due to ICME impact on Earth's
magnetosphere. The remaining 10\% of the large storms are caused by
corotating interaction regions (CIRs) resulting from the interaction
between fast and slow solar wind streams (see e.g., Zhang et~al.,
2007).

\subsection{Shock-driving CMEs}

CMEs driving fast mode MHD shocks can be directly observed in the
solar wind. Occurrence of type II radio bursts at the local plasma
frequency in the vicinity of the observing spacecraft (Bale et~al.,
1999) is strong evidence that the radio bursts are produced by
electrons accelerated at the shock front by the plasma emission
mechanism first proposed by Ginzburg and Zhelezniakov (1958).  The
frequency of type II burst emission is related to the plasma density
in the corona, so high frequency (about 150~MHz) type II bursts are
indicative of shocks accelerating electrons near the Sun. CMEs must
have speeds exceeding the local fast-mode speed in order to drive a
shock.  CMEs associated with metric type II bursts have a speed of
about 600~km/s, while those producing type II bursts at
decameter-hectometric (DH) wavelengths have an average speed exceeding
1100~km/s. Type II bursts with emission components from metric to
kilometric wavelengths are associated with the fastest CMEs (average
speed about $1500$~km/s).  Thus type II bursts are good indicators of
shock-driving CMEs (Gopalswamy et~al., 2005). Here we take type II
bursts at DH wavelengths to be indicative of CMEs driving IP shocks.
However, not all DH type II bursts are indicative of shocks detected
in situ.  This is mainly because CMEs originating even behind the
limbs can produce type II bursts because of the extended nature of the
shock and the wide beams of the radio emission, but these shocks need
not reach Earth.  On the other hand, there are some shocks detected at
1 AU, which are not associated with DH type II bursts; shocks have to
be of certain threshold strength to accelerate electrons.

A subset of shock-driving CMEs are associated with solar energetic
particle (SEP) events detected near Earth.  Naturally, the associated
CMEs form a subset of those producing DH type II bursts because the
same shocks accelerate electrons and ions.  In fact, all the major SEP
events are associated with DH type II bursts (Gopalswamy, 2003; Cliver
et~al., 2004), but only about half of the DH type II bursts have SEP
association (Gopalswamy et~al., 2008). CMEs associated with SEP events
have the highest average speed (about 1600~km/s).

\begin{table}[tb]
  \centering
\caption{Speed and width of the special populations of CMEs}
\label{gopalswamy-table:special} 
\begin{tabular}{lccccccc}
\hline
          & Halos & MCs & Non-MCs & Type IIs    & Shocks & Storms & SEPs \\ 
\hline
Speed (km/s)     & 1089 & 782 & 955 & 1194 & 966 & 1007 & 1557 \\ 
\% Halos         & 100  & 59  &  60 &   59 &  54 &   67 &   69 \\
\% Partial Halos & --   & 88  &  90 &   81 &  90 &   91 &   88 \\
Non-halo Width ($^\circ$)& --   & 55  &  84 &   83 &  90 &   89 &   48 \\ \hline
\end{tabular}
\end{table}

\subsection{Comparing the properties of the special populations}

\begin{figure}  [t]
  \centering
  \includegraphics[width=\textwidth]{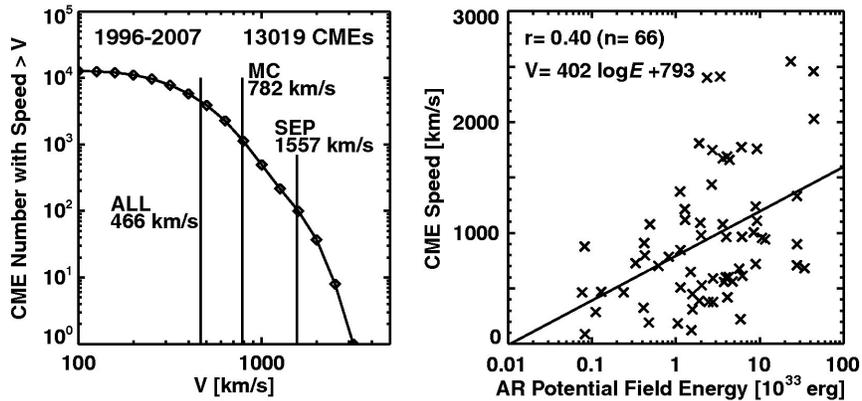}
  \caption[]{\label{gopalswamy-fig:cumulative}  
%
  {\em Left\/}: the cumulative number distribution of CMEs as a function of
  speed ($V$).  The average speed of all CMEs (466~km/s), CMEs
  resulting in MCs (782~km/s), and CMEs producing SEPs (1557~km/s) are
  indicated. Note that the number of CMEs with speeds $>$3000~km/s is
  negligibly small. {\em Right\/}: scatter plot between active-region (AR)
  potential energy ($E$) and $V$ for a set of 66 CMEs that had significant
  impact on Earth. These CMEs either produced large geomagnetic storms
  or became MCs. E was calculated using the AR flux ($\phi$) and area
  $A$ measured when the AR was close to the central meridian (CMD
  $\le$45$^\circ$) as follows: $E = \phi^2/8\pi\sqrt A$. The scatter
  plot shows a weak correlation ($r = 0.40$). The regression line is
  also shown on the plot.
}\end{figure}

Table~\ref{gopalswamy-table:special} compares the speed and width
information of the special population of CMEs discussed above.  The
lowest average speed is for MC-associated CMEs and the highest speed
is for SEP-producing CMEs.  The cumulative speed distribution of all
CMEs is shown in Fig.~\ref{gopalswamy-fig:cumulative} with the lowest
and highest speeds in Table~\ref{gopalswamy-table:special}
marked. Even the lowest speed (782 km/s for MCs) in
Table~\ref{gopalswamy-table:special} is well above the average speed
of all CMEs.  The average speed of the SEP-producing CMEs is the
highest (1557~km/s). All the other special populations have their
average speeds between these two limits.

The fraction of halo CMEs in a given population is an indicator of the
energy of the CMEs, because halo CMEs are more energetic on the
average owing to their higher speed and larger width. The majority of
CMEs in all special populations are halos. If partial halos are
included, the fraction becomes more than 80\% in each population. Even
the small fraction of non-halo CMEs ($W<120^\circ$) have an
above-average width.  The large fraction of halos in each population
implies that there is a high degree of overlap among the
populations, i.e., the same CME appears in various subgroups.

From Fig.~\ref{gopalswamy-fig:cumulative} one can see that the number
of CMEs with speeds $>$ 2000~km/s is exceedingly small. In fact, only
two CMEs are known to have speeds exceeding 3000 km/s among the more
than 13000 CMEs detected by SOHO during 1996 to 2007. This implies a
limit to the speed that CMEs can attain of about $4000$~km/s.  For a mass
of about $10^{17}$~g, a 4000~km/s CME would possess a kinetic energy of
$10^{34}$~erg. Active regions that produce such high energy CMEs
must possess a free energy of at least $10^{34}$~erg to power the
CMEs.  It has been estimated that the free energy in active regions is
of the order of the potential field energy and that the total magnetic
energy in the active region is about twice the potential field energy
(Mackay et~al., 1997; Metcalf et~al., 1995; Forbes, 2000;
Venkatakrishnan and Ravindra, 2003).  The potential field energy
depends on the size and the average magnetic field strength in the
active regions. Figure~\ref{gopalswamy-fig:cumulative} shows a scatter
plot between the potential field energy and the CME speed for a set of
CMEs that produced large geomagnetic storms or arrived at Earth as
MCs. The correlation is weak but there is certainly a trend that
faster CMEs arise from regions of higher potential field energy, as
previously shown by Venkatakrishnan and Ravindra (2003).  The limiting
speed of the CMEs can thus be traced to the maximum energy that can be
stored in solar active regions. The largest reported active region
size is about $5000$ millionths of a solar hemisphere (Newton, 1955),
and the highest magnetic field strength observed in sunspots is
about 6100~G (Livingston et~al., 2006). Combining these two, one can
estimate an upper limit of $10^{36}$~erg for the potential field
energy and hence the free energy that can be stored in an active
region. Usually a single CME does not exhaust all the free energy in
the active region. Note that some stars have much larger spots and can
release a lot more energy than in solar eruptions (Byrne, 1988).

\begin{figure}  
  \centering
  \includegraphics[width=\textwidth]{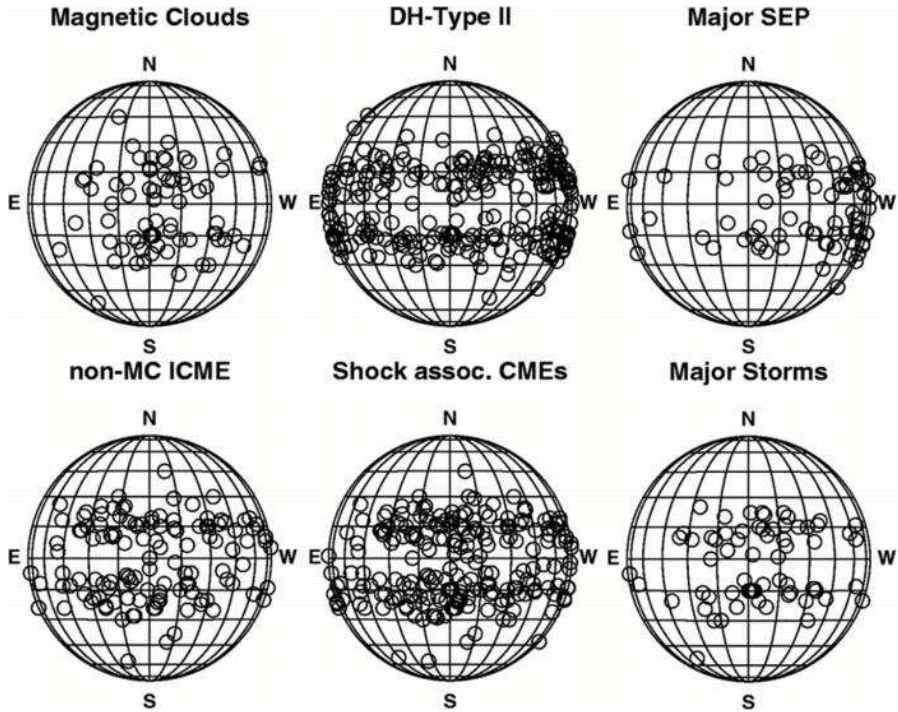}
  \caption[]{\label{gopalswamy-fig:gridplot}  
%
  Heliographic coordinates of the solar sources of the special
  populations.
}\end{figure}

\subsection{Solar sources of the special populations}

The lowest speed of the MC-associated CMEs and the highest speed of
the SEP-associated CMEs shown in Table~\ref{gopalswamy-table:special}
may be partly due to projection effects because the speeds were
measured in the sky plane. To see this, we have shown the
distributions of the solar sources of the special populations in
Fig.~\ref{gopalswamy-fig:gridplot}. The solar sources are taken as the
heliographic coordinates of the associated H$\alpha$ flares from the
Solar Geophysical Data. For events with no reported flare information,
we have taken the centroid of the post eruption arcade from EUV, X-ray
or microwave images as the solar source.  CMEs associated with MCs
generally originate from the disk center, so they are subject to
projection effects; the SEP-associated CMEs are mostly near the limb,
so the projection effects are expected to be minimal. Note that the
speed difference between MC- and SEP-associated CMEs is similar to
that of disk and limb halo CMEs (933 km/s vs.\ 1548 km/s -- see
Gopalswamy et~al., 2007). It is also possible that the SEP associated
CMEs are the fastest because they have to drive shocks and accelerate
particles.

The solar source distributions in Fig.~\ref{gopalswamy-fig:gridplot}
reveal several interesting facts: (i) Most of the sources are at low
latitudes with only a few exceptions during the rise phase.  (ii) The
MC sources are generally confined to the disk center, but the
non-cloud ICME sources are distributed at larger CMD. There is some
concentration of the non-MC sources to the east of the central
meridian. (iii) Subsets of MCs and non-MC ICMEs are responsible for
the major geomagnetic storms, so the solar sources of storm-associated
CMEs are also generally close to the central meridian. The slight
higher longitudinal extent compared to that of MC sources is due to
the fact that some storms are produced by shock sheaths of some fast
CMEs originating at larger CMD. (iv) The solar sources of CMEs
producing DH type II bursts have nearly uniform distribution in
longitude, including the east and west limbs. There are also sources
behind the east and west limbs that are not plotted.  The radio
emission can reach the observer from large angles owing to the wide
beam of the radio bursts. (v) The sources of SEP-associated CMEs, on
the other hand, are confined mostly to the western hemisphere with a
large number of sources close to the limb. In fact there are also many
sources behind the west limb, not plotted here (see Gopalswamy et~al.,
2008a for more details).  This western bias is known to be due to the
spiral structure of the IP magnetic field along which the SEPs have to
propagate before being detected by an observer near Earth. Typically
the longitude W70 is well connected to an Earth observer. An observer
located to the east is expected to detect more particle events from
the CMEs that produce DH type II bursts but located on the eastern
hemisphere. There are a few eastern sources producing SEPs, but these
are generally low-intensity events from very fast CMEs. (vi) The shock
sources are quite similar to the DH type II sources except for the
limb part. Since the associated CMEs need to produce a shock signature
at Earth, they are somewhat restricted to the disk. Occasional limb
CMEs did produce shock signatures at Earth, but these are shock
flanks. Comparison with DH type II sources reveals that many shocks do
not produce radio emission probably due to the low Mach number
(Gopalswamy et~al., 2008b).

It is also interesting to note that the combined MC and non-cloud ICME
source distribution is similar to those of halo CMEs and the ones
associated with shocks at 1~AU.  Even the sources of the SEP
associated CMEs are similar to the halos originating from the western
hemisphere of the Sun because of the requirement of magnetic
connectivity to the particle detector.

\begin{figure}  
  \centering
  \includegraphics[width=\textwidth]{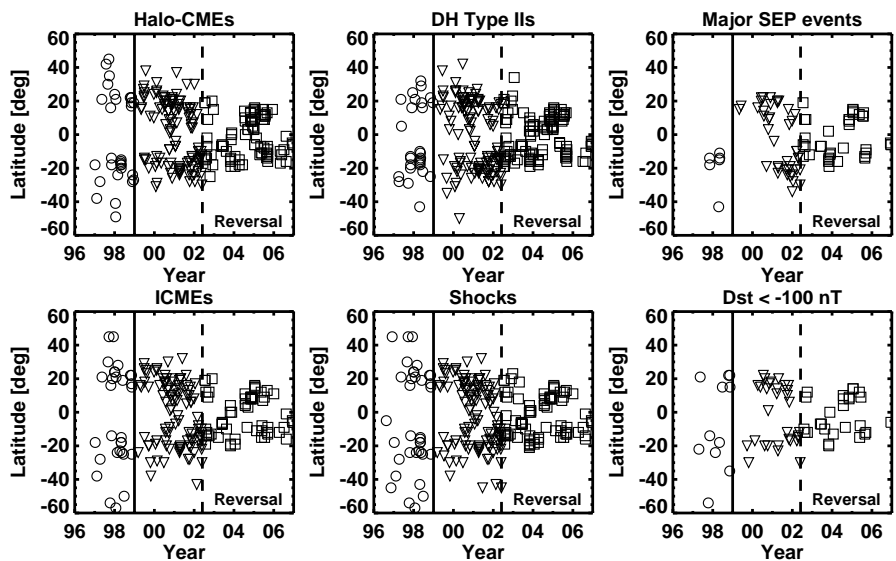}
  \caption[]{\label{gopalswamy-fig:cyclevlat}  
%
  The source latitudes of the special populations plotted as a
  function of time: halo CMEs, CMEs detected in situ as ICMEs, CMEs
  producing DH type II bursts, CMEs driving IP shocks, CMEs producing
  large SEP events, and CMEs resulting in large geomagnetic storms
  ($Dst\le-100$~nT). The solid vertical line (end of the year 1998)
  divides the rise and maximum phases. The dashed vertical line (end
  of May 2002) divides the maximum and declining phases. This line
  also marks the end of polarity reversal at solar poles (i.e.,
  beginning of the new solar magnetic cycle). CMEs from the three
  phases of solar cycle 23 are also distinguished by different symbols
  (rise - open circles, maximum - triangles, and declining - squares).
}\end{figure}

\section{Solar cycle variation}

CMEs originating close to the disk center and in the western
hemisphere have important implications to the space environment of
Earth because of the geomagnetic storms and the SEP events they
produce. Source regions of CMEs come close to the disk center in two
ways: (i) the solar rotation brings active regions to the central
meridian, and (ii) the progressive decrease in the latitudes where
active regions emerge from beneath the photosphere (the butterfly
diagram).  The effect due to the solar rotation is of short-term
because an active region stays in the vicinity of the disk center only
for 3--4 days during its disk passage. In order to see the effect of
the butterfly diagram, we need to plot the solar sources of as a
function of time.

Figure~\ref{gopalswamy-fig:cyclevlat} shows the latitude distribution
of the solar sources of the special populations as a function of time
during solar cycle 23. Sources corresponding to the three phases of
the solar cycle are distinguished using different symbols: the rise
phase starts from the beginning of the cycle in 1996 to the end of
1998. The maximum phase is taken from the beginning of 1999 to the
middle of 2002. The time of completion of the polarity reversal of the
solar polar magnetic fields is considered as the end of the solar
maximum phase and the beginning of the declining phase. The boundary
between phases is not precise, but one can see the difference in the
levels of activity and the change in latitude between phases.  We have
combined the MCs and non-MC ICMEs into a single group as ICMEs. As we
noted before, there is a close similarity between the ICME and halo
CME sources because many frontside halo CMEs become ICMEs (see the
left column in Fig.~\ref{gopalswamy-fig:cyclevlat}. There are clearly
ICMEs without corresponding halos, which means some ICMEs are due to
non-halo CMEs. During the rise phase there are several sources at
latitudes higher than the ones at which sunspots emerge
(about $40^\circ$). What is striking is that such high-latitude CMEs
produced an observable signature at Earth. Case studies (Gopalswamy
et~al., 2000, Gopalswamy \& Thompson, 2000) and statistical studies
(Plunkett et~al., 2001; Gopalswamy et~al., 2003b; Cremades et~al.,
2006) have shown that CMEs during the solar minimum get deflected
towards the equator because of the strong global dipolar field of the
Sun.  In the declining phase, a different type of deflection occurs:
eruptions occurring near low latitude coronal holes tend to be
deflected away from coronal holes. Occasionally, such deflections push
CMEs toward or away from the Sun-Earth line (Gopalswamy et~al.,
2009c).

\begin{figure}  [tb]
  \centering
  \includegraphics[width=\textwidth]{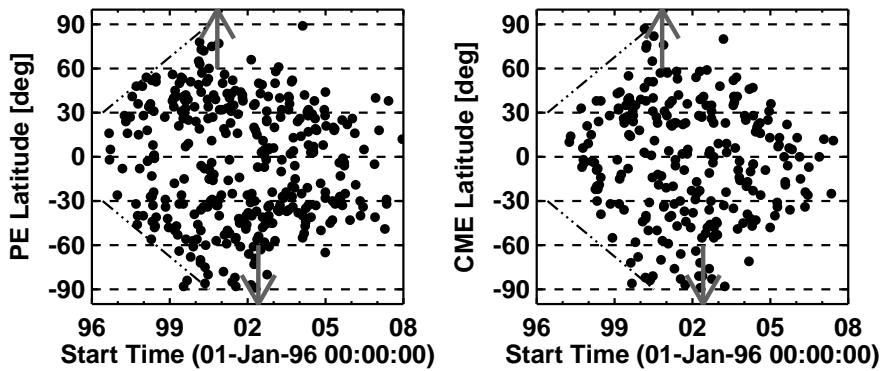}
  \caption[]{\label{gopalswamy-fig:norhpe}  
%
Latitude of prominence eruptions (PEs) and those of the associated
CMEs shown as a function of time.  The up and down arrows denote
respectively the times when the polarity in the north and south solar
poles reversed.  Note that the high-latitude CMEs and PEs are confined
to the solar maximum phase and their occurrence is asymmetric in the
northern and southern hemispheres.  PEs at latitudes below 40$^\circ$
may be from active regions or quiescent filament regions, but those at
higher latitudes are always from the latter.
}\end{figure}

\begin{figure}  
  \centering
  \includegraphics[width=\textwidth]{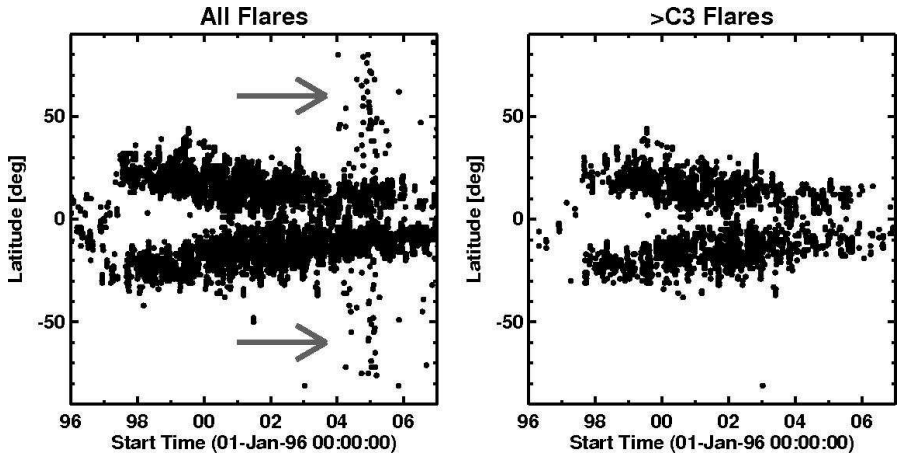}
  \caption[]{\label{gopalswamy-fig:flarelat}  
%
Flare locations reported in the Solar Geophysical Data plotted as a
function of time for all flares ({\em left\/}) and larger flares (soft X-ray
importance $>$C3.0) ({\em right\/}). The arrows point to the weak flares from
higher latitudes. Note that the GOES Soft X-ray imager provided solar
source locations of flares only during January 2004 to March 2007, so
there is no information on the high-latitude flare locations for other
times.
}\end{figure}

Despite the longitudinal differences presented in
Fig.~\ref{gopalswamy-fig:cyclevlat}, we see a close latitudinal
similarity of the solar sources of the special populations. The
sources clearly follow the sunspot butterfly diagram.  This means the
special populations originate only from active regions, where one
expects to have higher free energy needed to power these energetic
CMEs.

\subsection{Solar sources of the general population}

In contrast to the solar sources of the special populations discussed
above, the general CME population is known to occur at all latitudes
during solar maxima (Hundhausen, 1993; Gopalswamy et~al., 2003b).
Figure~\ref{gopalswamy-fig:norhpe} illustrates this using the latitude
distributions of prominence eruptions (PEs) and the associated CMEs.
Note that these CMEs constitute a very small sample because they are
chosen based on their association with PEs detected by the Nobeyama
radioheliograph (Nakajima et~al., 1994), which is a ground based
instrument operating only about $8$~h per day. Nevertheless, the
observations provide accurate source information for the CMEs and the
sample is not subject to projection effects.  One can clearly see a
large number of high latitude CMEs between the years 1999 and 2003,
with a significant north-south asymmetry in the source distributions.
These high-latitude CMEs are associated with polar crown filaments,
which migrate toward the solar poles and completely disappear by the
end of the solar maximum.  The cessation of high-latitude CME activity
has been found to be a good indicator of the polarity reversal at
solar poles (Gopalswamy et~al., 2003c).  Low-latitude PEs may be
associated with both active regions and quiescent filament regions,
but the high-latitude CMEs are always associated with filament
regions.  One can clearly see that the high-latitude CMEs have no
relation to the sunspot activity because the latter is confined to
latitudes below $40^\circ$.  Comparing
Figures~\ref{gopalswamy-fig:cyclevlat} and \ref{gopalswamy-fig:norhpe}
we can conclude that the special populations are primarily an active
region phenomenon. It is interesting that the high-latitude CMEs occur
only during the period of maximum sunspot number (SSN), but are not
directly related to the sunspots.

\subsection{Implications to the flare -- CME connection}

The difference in the latitude distributions of CMEs (no butterfly
diagram) and flares (follow the sunspot butterfly diagram) coupled with
the weak correlation between CME kinetic energy and soft X-ray flare
size (Hundhausen, 1997) has been suggested as evidence that CMEs are
not directly related to flares.  However, this depends on the
definition of flares. If flares are defined as the enhanced
electromagnetic emission from the structures left behind after CME
eruptions, one can find flares associated with all CMEs -- both at high
and at low latitudes. This is illustrated using
Fig.~\ref{gopalswamy-fig:flarelat}, which shows the solar source
locations of flares reported in the Solar Geophysical Data. During
2004 January - 2007 March, the GOES Soft X-ray Imager (SXI) provided
the solar sources of all flares, including the weak ones that can be
found at all latitudes, similar to the source distribution shown in
Fig.~\ref{gopalswamy-fig:norhpe} for PEs. On the other hand if we
consider only larger flares (X-ray importance $>$C3.0), we see that
the flares follow the sunspot butterfly diagram.  This is quite
consistent with the fact that the solar sources of the special
populations of CMEs follow the sunspot butterfly diagram because these
CMEs are associated with larger flares. For example, the median size
of flares associated with halo CMEs is M2.5, an order of magnitude
larger than the median size of all flares (C1.7) during solar cycle 23
(Gopalswamy et~al., 2007). Thus, CMEs seem to be related to flares
irrespective of the origin in active regions or quiescent filament
regions.  There are in fact several new indicators of the close
connection between CMEs and flares: CME speed and flare profiles
(Zhang et~al., 2001), CME and flare angular widths (Moore et~al.,
2007), CME magnetic flux in the IP medium and the reconnection flux at
the Sun (Qiu et~al., 2007), and the CME and flare positional
correspondence (Yashiro et~al., 2008a). The close relationship between
flares and CMEs does not contradict the fact that more than half of
the flares are not associated with CMEs. This is because the stored
energy in the solar source regions can be released to heat the flaring
loops with no mass motion.

\begin{figure}  
  \centering
  \includegraphics[width=\textwidth]{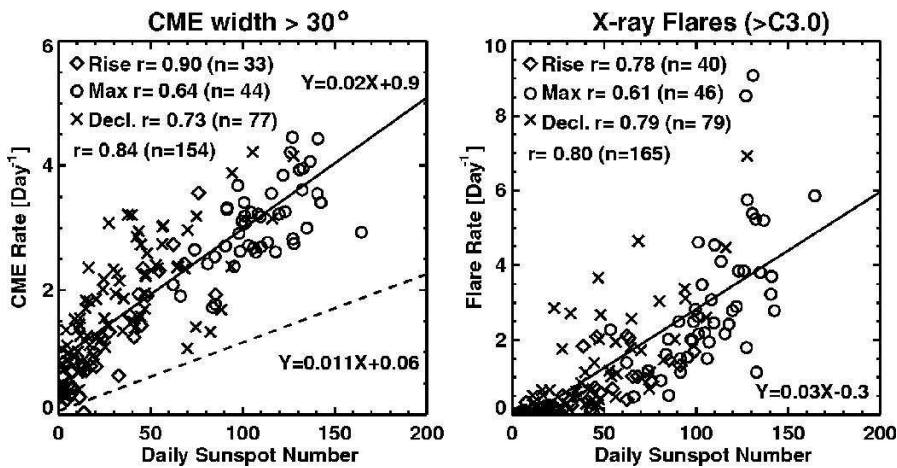}
  \caption[]{\label{gopalswamy-fig:ssn}  
%
Correlation of the daily sunspot number with the daily CME rate ({\em
left\/}) and the daily flare rate ({\em right\/}). All numbers are
averaged over Carrington rotation periods (27.3 days). The number of
rotations (n) is different for the CME and flare rates because of CME
data gaps. Rates in different phases of the solar cycle are shown by
different symbols. The correlation coefficients are also shown for
individual phases as well as for the entire data set (the solid lines
are the regression lines). In the CME rate, only CMEs wider than
30$^\circ$ are used to avoid subjectivity in CME identification.  In
the CME plot, the dashed line corresponds to the regression line (Y =
0.011X + 0.06) obtained by Cliver et~al. (1994) for CMEs from the
pre-SOHO era. In the flare rate, only flares of importance $>$C3.0 are
included.

}\end{figure}

\section{Sunspot number and CME rate}

The above discussion made it clear that the high-latitude CMEs do not
follow the sunspot butterfly diagram but occur during the period of
maximum solar activity.  This should somehow be reflected in the
relation between CME and sunspot activities. To see this, we have
plotted the daily CME rate ($R$) as a function of the daily sunspot
number (SSN) in Fig.~\ref{gopalswamy-fig:ssn}. There is an overall
good correlation between the two types of activity, which has been
known for a long time (Hildner et~al., 1976; Webb et~al. 1994; Cliver
et~al., 1994; Gopalswamy et~al., 2003a).  The SOHO data yield a
relation $R = 0.02\,\rm{SSN} + 0.9$ (correlation coefficient $r = 0.84$),
which has a larger slope compared to the one obtained by Cliver
et~al. (1994): $R = 0.011\,\rm{SSN} + 0.06$. The higher rate has been
attributed to the better dynamic range and wider field of view of the
SOHO coronagraphs compared to the pre-SOHO coronagraphs. However, when
the CMEs are grouped according to the phase of the solar cycle, the
correlation becomes weak during the maximum phase ($r=0.64$) compared
to the rise ($r=0.90$) and declining ($r=0.73$) phases. We attribute
this diminished correlation to the CMEs during the maximum phase that
are not associated with sunspots (see also Gopalswamy et~al., 2003a).
Note that we have excluded narrow CMEs ($W<30^\circ$) because manual
detection of such CMEs is highly subjective.

A similar scatter plot involving the daily flare rate as a function of
SSN reveals a similar trend in terms of the overall correlations
($r=0.80$) and the individual phases: rise ($r=0.78$), maximum
($r=0.61$) and declining ($r=0.79$).  In particular, the weak
correlation between the flare rate and SSN during the maximum phase is
striking. When all the flares are included, the overall correlation
diminished only slightly ($r=0.76$) mainly due to the weaker
correlation during the maximum phase ($r=0.46$) because the
correlation remained high during the rise ($r=0.85$), and declining
($r=0.79$) phases. Interestingly, flare rate vs.\ SSN correlations are
very similar to the CME rate vs.\ SSN correlations, including the weaker
correlation during the maximum phase. This needs further investigation
by separating the flares into high and low-latitude events.

\section{Summary and conclusions}

In this paper, we studied several subsets of CMEs that have
significant consequences in the heliosphere: halo CMEs, SEP-producing
CMEs, CMEs associated with IP type II radio bursts, CMEs associated
with shocks detected in situ in the solar wind, CMEs detected at 1 AU
as magnetic clouds and non-cloud ICMEs. The primary common property of
these special populations is their above-average energy, which helps
them propagate far into the IP medium.  Most of the CMEs in these
subsets are frontside halo CMEs. Notable exceptions are the IP type II
bursts and large SEP events. IP type II bursts can be observed from
CMEs from behind the east and west limbs because the shocks
responsible for the radio emission are more extended than the driving
CMEs, and the radio emission is wide beamed. SEP events are also
observed from behind the west limb for the same reason (extended
shock) and the fact that the SEPs propagate from the shock to the
observer along the spiral magnetic field lines. Another common
property of the special populations is that they follow the sunspot
butterfly diagram. This suggests that the energetic CMEs originate
mostly from the sunspot regions, where large free energy can be stored
to power the energetic CMEs.  Quiescent filament regions are the other
source of CMEs, not related to the sunspots, and hence do not follow
the sunspot butterfly diagram. During the maximum phase of the solar
activity cycle, the quiescent filament regions occur in high abundance
at high latitudes, resulting in higher rate of CMEs from there.  Since
the high-latitude CMEs are not related to the sunspots, the
correlation between daily CME rate and sunspot number is weak in the
maximum phase.  A similar weak correlation was found between the
flaring rate and sunspot number during the maximum phase. This result
further confirms the close relation between flares and CMEs
irrespective of the source region: sunspot regions or quiescent
filament regions.

\begin{acknowledgement}
 This work is supported by NASA's LWS program.
\end{acknowledgement}

\begin{small}

\end{small}

\end{document}

%% file: rr-assp-defs.tex

\def\thisvolume{these proceedings}

\def\aj{{AJ}}			
\def\araa{{ARA\&A}}		
\def\apj{{ApJ}}			
\def\apjl{{ApJ}}		
\def\apjs{{ApJS}}		
\def\ao{{Appl.\ Optics}} 
\def\apss{{Ap\&SS}}		
\def\aap{{A\&A}}		
\def\aapr{{A\&A~Rev.}}		
\def\aaps{{A\&AS}}		
\def\an{{Astron.\ Nachrichten}}
\def\aspcs{{ASP Conf.\ Ser.}}
\def\assp{{Astrophys.\ \& Space Sci.\ Procs., Springer, Heidelberg}}
\def\azh{{AZh}}			
\def\baas{{BAAS}}		
\def\jrasc{{JRASC}}	
\def\memras{{MmRAS}}		
\def\mnras{{MNRAS}}
\def\nat{{Nat}}		
\def\pra{{Phys.\ Rev.\ A}} 
\def\prb{{Phys.\ Rev.\ B}}		
\def\prc{{Phys.\ Rev.\ C}}		
\def\prd{{Phys.\ Rev.\ D}}		
\def\prl{{Phys.\ Rev.\ Lett.}} 
\def\pasp{{PASP}}
\def\pasj{{PASJ}}		
\def\qjras{{QJRAS}}
\def\science{{Sci}}		
\def\skytel{{S\&T}}		
\def\solphys{{Solar\ Phys.}} 
\def\sovast{{Soviet\ Ast.}}  
\def\ssr{{Space\ Sci.\ Rev.}}
\def\svassp{{Astrophys.\ Space Sci.\ Procs., Springer, Heidelberg}}
\def\zap{{ZAp}}			
\let\astap=\aap
\let\apjlett=\apjl
\let\apjsupp=\apjs
\def\grl{{Geophys.\ Res.\ Lett.}}  
\def\jgr{{J. Geophys.\ Res.}} 

\def\ion#1#2{{\rm #1}\,{\uppercase{#2}}}  
\def\deg{\hbox{$^\circ$}}
\def\sun{\hbox{$\odot$}}
\def\earth{\hbox{$\oplus$}}
\def\la{\mathrel{\hbox{\rlap{\hbox{\lower4pt\hbox{$\sim$}}}\hbox{$<$}}}}
\def\ga{\mathrel{\hbox{\rlap{\hbox{\lower4pt\hbox{$\sim$}}}\hbox{$>$}}}}
\def\sq{\hbox{\rlap{$\sqcap$}$\sqcup$}}
\def\arcmin{\hbox{$^\prime$}}
\def\arcsec{\hbox{$^{\prime\prime}$}}
\def\fd{\hbox{$.\!\!^{\rm d}$}}
\def\fh{\hbox{$.\!\!^{\rm h}$}}
\def\fm{\hbox{$.\!\!^{\rm m}$}}
\def\fs{\hbox{$.\!\!^{\rm s}$}}
\def\fdg{\hbox{$.\!\!^\circ$}}
\def\farcm{\hbox{$.\mkern-4mu^\prime$}}
\def\farcs{\hbox{$.\!\!^{\prime\prime}$}}
\def\fp{\hbox{$.\!\!^{\scriptscriptstyle\rm p}$}}
\def\micron{\hbox{$\mu$m}}
\def\onehalf{\hbox{$\,^1\!/_2$}}	
\def\onethird{\hbox{$\,^1\!/_3$}}
\def\twothirds{\hbox{$\,^2\!/_3$}}
\def\onequarter{\hbox{$\,^1\!/_4$}}
\def\threequarters{\hbox{$\,^3\!/_4$}}
\def\ubv{\hbox{$U\!BV$}}		
\def\ubvr{\hbox{$U\!BV\!R$}}		
\def\ubvri{\hbox{$U\!BV\!RI$}}		
\def\ubvrij{\hbox{$U\!BV\!RI\!J$}}		
\def\ubvrijh{\hbox{$U\!BV\!RI\!J\!H$}}		
\def\ubvrijhk{\hbox{$U\!BV\!RI\!J\!H\!K$}}		
\def\ub{\hbox{$U\!-\!B$}}		
\def\bv{\hbox{$B\!-\!V$}}		
\def\vr{\hbox{$V\!-\!R$}}		
\def\ur{\hbox{$U\!-\!R$}}


\def\labelitemi{{\bf --}}  

\def\rmit#1{{\it #1}}              
\def\rmit#1{{\rm #1}}              
\def\etal{\rmit{et al.}}           
\def\etc{\rmit{etc.}}           
\def\ie{\rmit{i.e.,}}              
\def\eg{\rmit{e.g.,}}              
\def\cf{cf.}                       
\def\viz{\rmit{viz.}}
\def\vs{\rmit{vs.}}

\def\rot{\hbox{\rm rot}}
\def\div{\hbox{\rm div}}
\def\lesssim{\mathrel{\hbox{\rlap{\hbox{\lower4pt\hbox{$\sim$}}}\hbox{$<$}}}}
\def\gtrsim{\mathrel{\hbox{\rlap{\hbox{\lower4pt\hbox{$\sim$}}}\hbox{$>$}}}}
\def\dif{\: {\rm d}}                       
\def\ep{\:{\rm e}^}                        
\def\dash{\hbox{$\,-\,$}}                  
\def\is{\!=\!}                             

\def\starname#1#2{${#1}$\,{\rm {#2}}}  
\def\Teff{\hbox{$T_{\rm eff}$}}   

\def\kms{\hbox{km$\;$s$^{-1}$}}
\def\ms{\hbox{m$\;$s$^{-1}$}}
\def\Mxcm{\hbox{Mx\,cm$^{-2}$}}    

\def\Bapp{\hbox{$B_{\rm app}$}}    

\def\komega{($k, \omega$)}                 
\def\kf{($k_h,f$)}                         
\def\VminI{\hbox{$V\!\!-\!\!I$}}           
\def\IminI{\hbox{$I\!\!-\!\!I$}}           
\def\VminV{\hbox{$V\!\!-\!\!V$}}           
\def\Xt{\hbox{$X\!\!-\!t$}}                

\def\level #1 #2#3#4{$#1 \: ^{#2} \mbox{#3} ^{#4}$}   

\def\specchar#1{\uppercase{#1}}    
\def\AlI{\mbox{Al\,\specchar{i}}}  
\def\BI{\mbox{B\,\specchar{i}}} 
\def\BII{\mbox{B\,\specchar{ii}}}  
\def\BaI{\mbox{Ba\,\specchar{i}}}  
\def\BaII{\mbox{Ba\,\specchar{ii}}} 
\def\CI{\mbox{C\,\specchar{i}}} 
\def\CII{\mbox{C\,\specchar{ii}}} 
\def\CIII{\mbox{C\,\specchar{iii}}} 
\def\CIV{\mbox{C\,\specchar{iv}}} 
\def\CaI{\mbox{Ca\,\specchar{i}}} 
\def\CaII{\mbox{Ca\,\specchar{ii}}} 
\def\CaIII{\mbox{Ca\,\specchar{iii}}} 
\def\CoI{\mbox{Co\,\specchar{i}}} 
\def\CrI{\mbox{Cr\,\specchar{i}}} 
\def\CriI{\mbox{Cr\,\specchar{ii}}} 
\def\CsI{\mbox{Cs\,\specchar{i}}} 
\def\CsII{\mbox{Cs\,\specchar{ii}}} 
\def\CuI{\mbox{Cu\,\specchar{i}}} 
\def\FeI{\mbox{Fe\,\specchar{i}}} 
\def\FeII{\mbox{Fe\,\specchar{ii}}} 
\def\FeIX{\mbox{Fe\,\specchar{ix}}}
\def\FeX{\mbox{Fe\,\specchar{x}}}
\def\FeXVI{\mbox{Fe\,\specchar{xvi}}}
\def\FrI{\mbox{Fr\,\specchar{i}}}
\def\HI{\mbox{H\,\specchar{i}}} 
\def\HII{\mbox{H\,\specchar{ii}}} 
\def\Hmin{\hbox{\rmH$^{^{_{\scriptstyle -}}}$}}      
\def\Hemin{\hbox{{\rm He}$^{^{_{\scriptstyle -}}}$}} 
\def\HeI{\mbox{He\,\specchar{i}}} 
\def\HeII{\mbox{He\,\specchar{ii}}} 
\def\HeIII{\mbox{He\,\specchar{iii}}} 
\def\KI{\mbox{K\,\specchar{i}}} 
\def\KII{\mbox{K\,\specchar{ii}}} 
\def\KIII{\mbox{K\,\specchar{iii}}} 
\def\LiI{\mbox{Li\,\specchar{i}}} 
\def\LiII{\mbox{Li\,\specchar{ii}}} 
\def\LiIII{\mbox{Li\,\specchar{iii}}} 
\def\MgI{\mbox{Mg\,\specchar{i}}} 
\def\MgII{\mbox{Mg\,\specchar{ii}}} 
\def\MgIII{\mbox{Mg\,\specchar{iii}}} 
\def\MnI{\mbox{Mn\,\specchar{i}}} 
\def\NI{\mbox{N\,\specchar{i}}}
\def\NIV{\mbox{N\,\specchar{iv}}}
\def\NaI{\mbox{Na\,\specchar{i}}}
\def\NaII{\mbox{Na\,\specchar{ii}}}
\def\NaIII{\mbox{Na\,\specchar{iii}}}
\def\NeVIII{\mbox{Ne\,\specchar{viii}}} 
\def\NiI{\mbox{Ni\,\specchar{i}}} 
\def\NiII{\mbox{Ni\,\specchar{ii}}}
\def\NiIII{\mbox{Ni\,\specchar{iii}}} 
\def\OI{\mbox{O\,\specchar{i}}} 
\def\OVI{\mbox{O\,\specchar{vi}}}
\def\RbI{\mbox{Rb\,\specchar{i}}} 
\def\SII{\mbox{S\,\specchar{ii}}} 
\def\SiI{\mbox{Si\,\specchar{i}}} 
\def\SiII{\mbox{Si\,\specchar{ii}}} 
\def\SrI{\mbox{Sr\,\specchar{i}}}
\def\SrII{\mbox{Sr\,\specchar{ii}}}
\def\TiI{\mbox{Ti\,\specchar{i}}} 
\def\TiII{\mbox{Ti\,\specchar{ii}}} 
\def\TiIII{\mbox{Ti\,\specchar{iii}}} 
\def\TiIV{\mbox{Ti\,\specchar{iv}}} 
\def\VI{\mbox{V\,\specchar{i}}} 
\def\HtwoO{\mbox{H$_2$O}}        
\def\Otwo{\mbox{O$_2$}}          

\def\Halpha{\mbox{H\hspace{0.1ex}$\alpha$}} 
\def\Ha{\mbox{H\hspace{0.2ex}$\alpha$}}
\def\Hbeta{\mbox{H\hspace{0.2ex}$\beta$}}
\def\Hgamma{\mbox{H\hspace{0.2ex}$\gamma$}}
\def\Hdelta{\mbox{H\hspace{0.2ex}$\delta$}}
\def\Hepsilon{\mbox{H\hspace{0.2ex}$\epsilon$}}
\def\Hzeta{\mbox{H\hspace{0.2ex}$\zeta$}}
\def\Lyalpha{\mbox{Ly$\hspace{0.2ex}\alpha$}}
\def\Lybeta{\mbox{Ly$\hspace{0.2ex}\beta$}}
\def\Lygamma{\mbox{Ly$\hspace{0.2ex}\gamma$}}
\def\Lycont{\mbox{Ly\hspace{0.2ex}{\small cont}}}
\def\Baalpha{\mbox{Ba$\hspace{0.2ex}\alpha$}}
\def\Babeta{\mbox{Ba$\hspace{0.2ex}\beta$}}
\def\Bacont{\mbox{Ba\hspace{0.2ex}{\small cont}}}
\def\Paalpha{\mbox{Pa$\hspace{0.2ex}\alpha$}}
\def\Bralpha{\mbox{Br$\hspace{0.2ex}\alpha$}}

\def\NaD{\mbox{Na\,\specchar{i}\,D}}    
\def\NaDone{\mbox{Na\,\specchar{i}\,\,D$_1$}}
\def\NaDtwo{\mbox{Na\,\specchar{i}\,\,D$_2$}}
\def\NaID{\mbox{Na\,\specchar{i}\,\,D}}
\def\NaIDone{\mbox{Na\,\specchar{i}\,\,D$_1$}}
\def\NaIDtwo{\mbox{Na\,\specchar{i}\,\,D$_2$}}
\def\Done{\mbox{D$_1$}}
\def\Dtwo{\mbox{D$_2$}}

\def\Mgbone{\mbox{Mg\,\specchar{i}\,b$_1$}}
\def\Mgbtwo{\mbox{Mg\,\specchar{i}\,b$_2$}}
\def\Mgbthree{\mbox{Mg\,\specchar{i}\,b$_3$}}
\def\MgIb{\mbox{Mg\,\specchar{i}\,b}}
\def\MgIbone{\mbox{Mg\,\specchar{i}\,b$_1$}}
\def\MgIbtwo{\mbox{Mg\,\specchar{i}\,b$_2$}}
\def\MgIbthree{\mbox{Mg\,\specchar{i}\,b$_3$}}

\def\CaIIK{\mbox{Ca\,\specchar{ii}\,K}}       
\def\CaIIH{\mbox{Ca\,\specchar{ii}\,H}}
\def\CaIIHK{\mbox{Ca\,\specchar{ii}\,H\,\&\,K}}
\def\HK{\mbox{H\,\&\,K}}
\def\Kthree{\mbox{K$_3$}}      
\def\Hthree{\mbox{H$_3$}}
\def\Ktwo{\mbox{K$_2$}}
\def\Htwo{\mbox{H$_2$}}
\def\Kone{\mbox{K$_1$}}     
\def\Hone{\mbox{H$_1$}}     
\def\KtwoV{\mbox{K$_{2V}$}}
\def\KtwoR{\mbox{K$_{2R}$}}
\def\KoneV{\mbox{K$_{1V}$}}
\def\KoneR{\mbox{K$_{1R}$}}
\def\HtwoV{\mbox{H$_{2V}$}}
\def\HtwoR{\mbox{H$_{2R}$}}
\def\HoneV{\mbox{H$_{1V}$}}
\def\HoneR{\mbox{H$_{1R}$}}

\def\hk{\mbox{h\,\&\,k}}
\def\kthree{\mbox{k$_3$}}    
\def\hthree{\mbox{h$_3$}}
\def\ktwo{\mbox{k$_2$}}
\def\htwo{\mbox{h$_2$}}
\def\kone{\mbox{k$_1$}}     
\def\hone{\mbox{h$_1$}}     
\def\ktwoV{\mbox{k$_{2V}$}}
\def\ktwoR{\mbox{k$_{2R}$}}
\def\koneV{\mbox{k$_{1V}$}}
\def\koneR{\mbox{k$_{1R}$}}
\def\htwoV{\mbox{h$_{2V}$}}
\def\htwoR{\mbox{h$_{2R}$}}
\def\honeV{\mbox{h$_{1V}$}}
\def\honeR{\mbox{h$_{1R}$}}

\ifnum\preprintheader=1     
\makeatletter  
\def\@maketitle{\newpage
\markboth{}{}%
  {\mbox{} \vspace*{-8ex} \par 
   \em \footnotesize To appear in ``Magnetic Coupling between the Interior 
       and the Atmosphere of the Sun'', eds. S.~S.~Hasan and R.~J.~Rutten, 
       Astrophysics and Space Science Proceedings, Springer-Verlag, 
       Heidelberg, Berlin, 2009.} \vspace*{-5ex} \par
 \def\lastand{\ifnum\value{@inst}=2\relax
                 \unskip{} \andname\
              \else
                 \unskip \lastandname\
              \fi}%
 \def\and{\stepcounter{@auth}\relax
          \ifnum\value{@auth}=\value{@inst}%
             \lastand
          \else
             \unskip,
          \fi}%
  \raggedright
 {\Large \bfseries\boldmath
  \pretolerance=10000
  \let\\=\newline
  \raggedright
  \hyphenpenalty \@M
  \interlinepenalty \@M
  \if@numart
     \chap@hangfrom{}
  \else
     \chap@hangfrom{\thechapter\thechapterend\hskip\betweenumberspace}
  \fi
  \ignorespaces
  \@title \par}\vskip .8cm
\if!\@subtitle!\else {\large \bfseries\boldmath
  \vskip -.65cm
  \pretolerance=10000
  \@subtitle \par}\vskip .8cm\fi
 \setbox0=\vbox{\setcounter{@auth}{1}\def\and{\stepcounter{@auth}}%
 \def\thanks##1{}\@author}%
 \global\value{@inst}=\value{@auth}%
 \global\value{auco}=\value{@auth}%
 \setcounter{@auth}{1}%
{\lineskip .5em
\noindent\ignorespaces
\@author\vskip.35cm}
 {\small\institutename\par}
 \ifdim\pagetotal>157\p@
     \vskip 11\p@
 \else
     \@tempdima=168\p@\advance\@tempdima by-\pagetotal
     \vskip\@tempdima
 \fi
}
\makeatother     
\fi